# Transmission characteristics of millimeter and sub-terahertz channels through spatially ripple plasma sheath layers

Wenbo Liu[1,2], Peian Li[1,2], Da Li[1,2], Daniel M. Mittleman[3], Jianjun Ma[1,2]

[1]School of Integrated Circuits and Electronics, Beijing Institute of Technology, Beijing 100081 China

[2]Beijing Key Laboratory of Millimeter Wave and Terahertz Technology, Beijing 100081 China

[3]School of Engineering, Brown University, Providence, RI 02912 USA

**ABSTRACT**

 The propagation of millimeter wave (MMW) and sub-terahertz (THz) signals through plasma sheaths is a critical concern for maintaining communication with hypersonic vehicles, yet the impact of complex plasma structures on these high-frequency channels remains insufficiently understood. In this work, we aim to characterize the transmission properties of MMW and sub-THz waves through plasma sheaths with various density profiles and ripple structures, addressing the gap in knowledge regarding the effects of plasma inhomogeneities on signal propagation. We employ an approach combining Inductively Coupled Plasma (ICP) data with transfer matrix methods (TMM) to model propagation through both flat and rippled plasma layers. Our findings reveal that ripple structures in plasma sheaths significantly affect channel performance, with periodic ripples reducing cutoff frequency and introducing frequency-selective behavior, while random ripples cause more unpredictable transmission characteristics. Our results explore the impact of the arrangement of plasma density layers and the parameters of ripple structures (period and amplitude) on channel transmission, group velocity dispersion, and angular dependence of wave propagation. These results provide crucial insights for the design and optimization of communication systems for hypersonic vehicles, potentially enabling the development of adaptive technologies capable of maintaining reliable communication in complex plasma environments.

Index Terms— Millimeter and sub-terahertz channel, ripple plasma sheath layers, transmission characteristics, group velocity dispersion

## I. INTRODUCTION

Plasma environments significantly influence the propagation of wireless channels, a phenomenon that has garnered substantial research interest due to its wide-ranging implications. The unique transmission and reflection properties in plasma environments have been leveraged to develop various technologies, including plasma filters [1, 2], frequency-selective surfaces [3], and plasma stealth devices [4]. As space technology has advanced rapidly, the flight speeds of hypersonic aircraft have dramatically increased, now reaching 10-20 Mach [5, 6]. This increase in velocity has exacerbated the persistent issue of communication

"blackout" during re-entry or hypersonic flight. At these extreme speeds, intense atmospheric friction generates an exceedingly high-temperature, high-pressure environment around the aircraft, leading to substantial ionization of atmospheric gases. The resulting plasma sheath envelops the entire surface of the aircraft, with electron densities reaching up to $10^{13} cm^{-3}$ [7]. This plasma layer acts as a barrier to wireless channels, particularly those below the plasma frequency, creating a cutoff frequency that can extend to several tens of GHz. Consequently, conventional radio frequency (RF) channels are effectively blocked or severely attenuated. Overcoming these challenges to achieve real-time communication linkshas been a primary focus of research in this field. Since the 1960s, when the NASA "RAM" project began investigating radiotransmission through plasma media, various solutions have been proposed to mitigate the "blackout" phenomenon [8]. These approaches span both physics and chemistry, including techniques such as gas discharge [9], magnetic windows [10], electrophilic fluid injection [11], high-frequency communication [12] and aerodynamic shape modifications [13]. Despite these efforts, the communication blackout problem remains a significant challenge in hypersonic flight, necessitating continued research and innovative solutions.

The advancement of wireless communication technologies towards higher frequency bands, particularly in the millimeter wave (MMW) and sub-terahertz (sub-THz) ranges, has opened up new possibilities for high-capacity data transmission and wide bandwidth applications [14-16]. This shift to higher frequencies is driven by the increasing demands for enhanced data throughput, improved detection resolution, and more efficient utilization of the electromagnetic spectrum. A key advantage of these higher frequency bands lies in their interaction with plasma environments. When the operating frequency significantly surpasses the electron plasma frequency, the plasma electrons experience substantial inertia effects, resulting in weak oscillations at the signal frequency. This phenomenon enables wireless channels to penetrate plasma barriers, such as the re-entry plasma sheath, with minimal attenuation [8, 17]. Despite the engineering challenges and increased costs associated with developing and implementing high-frequency communication systems, their potential benefits in overcoming plasma-induced signal blockage have sparked growing interest in MMW and sub-THz technologies for aerospace applications. This trend highlights the critical importance of thoroughly understanding the propagation characteristics of these high-frequency channels within plasma environments, particularly through plasma sheaths.

Previous investigations into electromagnetic wave propagation through plasma environments have predominantly focused on idealized conditions, assuming uniform plasma distributions, stable environmental parameters, and minimal external interference. These studies typically employed flow field models or theoretical calculations, such as scattering matrices, to establish plasma parameter distributions and assess the impact of plasma sheaths on channel transmission [18-20]. Researchers have extensively analyzed phenomena including reflection, transmission, absorption, and phase shifts in these idealized plasma conditions [21-23]. However, the field has been constrained by the significant challenges associated with accurately simulating transition and turbulence in hyper-velocity flight scenarios. These challenges, recognized as critical issues in Computational Fluid Dynamics (CFD) [24, 25], have only recently begun to be addressed through more sophisticated simulation techniques [26-28]. The limitations in computational capabilities have resulted in a lack of comprehensive studies on the effects of inhomogeneous plasma density distributions and small-scale fluctuations within plasma sheaths on MMW channel transmission. This gap in research persists despite substantial evidence demonstrating the propagation capabilities of these high-frequency waves through plasma environments. The importance of these small-scale structural variations is particularly significant for MMW and sub-THz frequency bands, where the relatively smaller wavelengths make the channels more susceptible to scattering effects caused by plasma inhomogeneities [16]. Notably, recent advancements in CFD and plasma flow field distribution simulations have started to reveal these complex dynamics, showing ripple localized concentration fluctuations within plasma sheaths [26-29]. These findings underscore the need for a more nuanced understanding of plasma-channel interactions in realistic, non-ideal conditions to accurately predict and optimize the performance of high-frequency communication systems in hypersonic and re-entry scenarios.

In this work, we conduct a comprehensive investigation into the transmission properties of MMW and sub-THz channels through plasma sheaths exhibiting spatial concentration fluctuations (i.e., ripples). Our approach leverages characteristic plasma data derived from the Inductively Coupled Plasma (ICP) model to construct realistic ripple plasma structures. We develop propagation models based on the Fresnel equations and employ the transfer matrix method (TMM) for theoretical validation and computation. This methodology allows us to analyze in detail the impact of ripple layers and incident angle variations on the propagation characteristics of MMW and sub-THz channels. To further explore the propagation attributes in small-scale plasma sheaths, we utilize plasma aerodynamic simulations and concentration distribution data from NASA RAM-C hypersonic re-entry vehicle experiments [7, 8]. By constructing scale-comparable plasma ripple models, we examine the effects of both transition-induced and turbulence-induced plasma ripple formations on channel propagation across various scales. This approach enables us to bridge the gap between idealized plasma models and the complex, inhomogeneous plasma environments encountered in real-world hypersonic flight scenarios.

## II. CHARACTERISTICS OF PLASMA ARRANGEMENT

To establish a foundation for investigating channel transmission characteristics in controlled laboratory plasma environments, we employ data from Inductively Coupled Plasma (ICP) devices as our primary parameter set. Although this is a purely computational investigation with no experimental measurements, it is still valuable to use ICP data. ICP devices offer a well-characterized and controllable plasma environment that closely approximates key aspects of re-entry plasma sheaths. This similarity is crucial for developing accurate computational models that can reliably predict channel propagation behavior in actual hypersonic flight scenarios. The plasma density distributions and spatial arrangements generated by these widely used plasma sources exhibit notable similarities to those observed in the plasma sheaths surrounding re-entry vehicles along specific transmission paths [20, 29]. Secondly, the use of ICP data allows us to incorporate realistic plasma inhomogeneities and fluctuations into our models. These subtle variations are often overlooked in idealized plasma simulations but can significantly impact channel propagation, especially at MMW and sub-THz frequencies. By basing our computations on ICP-derived data, we can capture these differences and provide more accurate predictions of channel behavior. As illustrated in Fig. 1, the ICP-generated plasma structure typically follows a 'Low-High-Low' (L-H-L) concentration profile from the outer edges towards the center. This configuration closely mimics the density distribution found in real-world re-entry plasma sheaths. Importantly, these ICP devices are capable of producing electron densities reaching up to $10^{13} cm^{-3}$ [30], which aligns well with the density ranges encountered in actual hypersonic flight scenarios. Furthermore, the use of ICP-based models in our computational study serves as a bridge between theoretical plasma physics and practical aerospace applications. While direct measurements of plasma sheaths during hypersonic flight are extremely challenging and rare, ICP devices provide an accessible means to validate computational models against experimental data. This validation, though not directly performed in this study, lends credibility to our computational approach and enhances the potential for future experimental verification of our findings. By utilizing ICP data, we can create a controlled and reproducible plasma environment that effectively simulates the key characteristics of re-entry plasma sheaths. This approach enables more accurate and relevant studies of MMW and sub-THz channel propagation through these complex media, providing insights that are both theoretically sound and practically applicable to real-world hypersonic communication challenges.

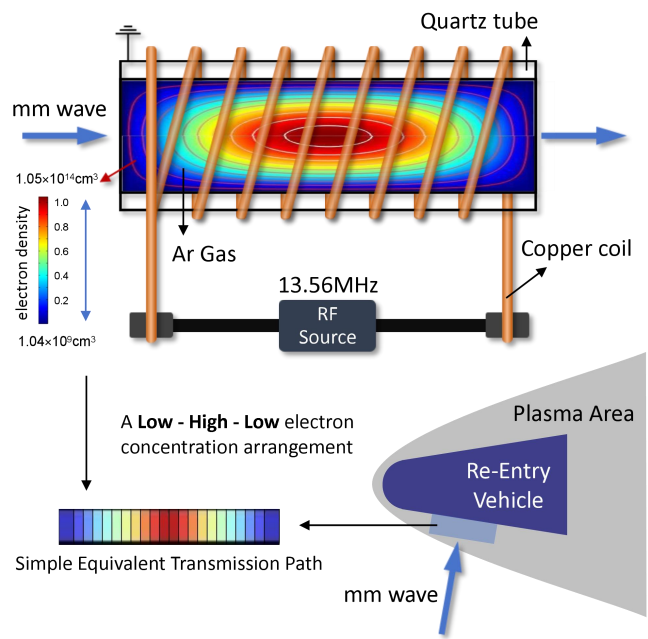

**Fig. 1.** Schematic diagram of the ICP device model and the plasma sheath surrounding a reentry vehicle.

*A. Plasma parameter simulation*

In our simulation model, we optimize computational efficiency by focusing on the essential components of the ICP system, incorporating five collision types and four chain reactions, as illustrated in Fig. 1 and Table 1. This model uses standard notation as "e" for electrons, "Ar" for ground-state argon, "Ar+" for argon ions, and "Ars" for excited-state argon.

TABLE I FIVE COLLISION TYPES & FOUR CHAIN REACTIONS

| Ar reaction | Type |
|---|---|
| Collision | e+Ar=>e+Ar <br> e+Ar=>e+Ars <br> e+Ars=>e+Ar <br> e+Ar=>2e+Ar+ <br> e+Ars=>2e+Ar+ |
| Chain reaction | 2Ars=>e+Ar+Ar+ <br> Ars+Ar=>2Ar <br> Ars=>Ar <br> Ars=>Ar |

The ICP source consists of two key components: an RF antenna and a sealed quartz tube, which generate and confine the quasi-neutral plasma. The antenna is a 5 mm diameter copper coil, wound in 9 turns around a cylindrical quartz tube (200 mm length, 100 mm diameter, 2 mm thickness). We operate the RF power source at 550 W and 13.56 MHz [31], with argon gas at 1.04 torr pressure, creating an RF glow discharge plasma. Our simulations show the ICP system stabilizes within 1 second, maintaining electron densities between $1.05 \times 10^{14}$ cm$^{-3}$ and $1.04 \times 10^{9}$ cm$^{-3}$, with density decreasing from the center outwards.

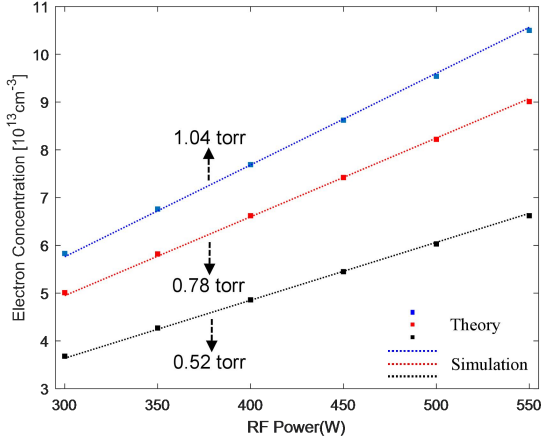

**Fig. 2.** Relationship between electron density, pressure, and RF power in the ICP apparatus.

To validate our model, we employ an ICP electromagnetic density calculation based on ambipolar diffusion [30, 32]

$$n_e = g \frac{P}{E/\Lambda^2} p^2 \quad (1)$$

Here $n_e$ is the electron density, $P$ is RF power of the container, $E$ is residual electric field, $p$ is gas pressure, $\Lambda$ is the inverse density gradient, and $g$ is a geometric parameter (set to 0.0122 based on similar ICP configurations [30]). At high plasma densities, $E/\Lambda^2$ may fall below 1, indicating increased uniformity, though this is challenging to measure directly [30][33].

Our simulations closely match theoretical predictions when $E/\Lambda^2$ equals 0.687, 0.45, and 0.272, corresponding to argon pressures of 1.04, 0.78, and 0.52 torr, respectively, as shown in Fig. 2. We observe a linear relationship between electron density and RF power at given pressures, with higher pressures yielding significantly higher electron densities. Based on these results, we select 550W RF power and 1.04 torr pressure for our subsequent simulations to achieve the high electron densities characteristic of plasma sheaths in hypersonic flight conditions.

*B. Simulation and theoretical calculation methods*

We conduct simulations using finite element analysis (FEA) and employ a narrow beam approach with a 2 cm beam-width, to mitigate the lens effect caused by plasma inhomogeneity as observed in [29], covering frequencies from 30 GHz to 300 GHz. Perform theoretical calculations using Fresnel equations to construct transmission matrices. This method allows us to approximate the beam coverage area as a series of discrete layers, each with uniform properties. While the actual plasma density varies continuously, we model it as a set of distinct layers with step changes in density between them. This layered approximation effectively creates a ripple plasma structure, where each layer represents a different density region of the plasma. This approach, while simplifying the continuous nature of plasma density variation, enables us to capture the essential features of plasma inhomogeneity and its effects on wave propagation [34, 35]. By adjusting the number and properties of these layers, we can simulate various degrees of plasma inhomogeneity, from smooth gradients to more abrupt changes in density. We also select the beam's central propagation axis as the data sampling line and reconstruct the electron density in layers. We define 22 layers in total, including air layers, with 10 representative plasma concentrations ranging from $1\times10^{13}$ cm$^{-3}$ to $10\times10^{13}$ cm$^{-3}$, incrementing by $1\times10^{13}$ cm$^{-3}$. This layered approach enables a comprehensive study of concentration distribution effects on channel transmission characteristics. The outermost layers are air (refractive index = 1), with electron density increasing towards the core, forming a 'Low-High-Low' (L-H-L) concentration arrangement, as shown in Fig. 1 and 3.

Given the significant inhomogeneity of the plasma sheath, accurate computation of dielectric parameters and refractive indices for each plasma layer is crucial. We derive these from ICP simulations and NASA re-entry vehicle research data [8], as replicating actual plasma sheath conditions in a laboratory is challenging. We approximate the plasma frequency $\omega_p$ by the electron plasma frequency $\omega_{pe}$, given that $\omega_{pe} >> \omega_{pi}$ (ion plasma frequency) and $m_e << m_i$ (electron and ion masses). The plasma frequency is calculated by the formula

$$\omega_p \approx \omega_{pe} = \sqrt{\frac{n_e e^2}{m_e \varepsilon_0}} \qquad (2)$$

where $e$ represents the elementary charge of an electron, and $\varepsilon_0$ is the vacuum permittivity. The plasma sheath permittivity is $\varepsilon = \varepsilon_0 \varepsilon_r$, with $\varepsilon_r$ being the relative permittivity expressed as $\varepsilon_r = 1 - \omega_p^2/(\omega^2 + v_{en}^2) - j v_{en} \cdot \omega_p^2 /[\omega \cdot (\omega^2 + v_{en}^2)]$. Here $\omega$ is the angular frequency of the incident electromagnetic wave, and $v_{en}$ is the plasma collision frequency. The plasma refractive index is then derived as $n = \sqrt{\varepsilon_r}$.

To analyze channel propagation characteristics from 30 GHz to 300 GHz, we combine Fresnel equations with the transfer matrix method (TMM) [36, 37]. Starting with an initial 2D unit transmission matrix $M_i$, we express the Fresnel coefficients for reflection ($r$) and transmission ($t$) at the interface between the first two layers under normal incidence as $r = (n_1 - n_2)/(n_1 + n_2)$ and $t = 2n_1/(n_1 + n_2)$, respectively. The propagation matrix after the second plasma layer can be derived as

$$M = M_i \frac{1}{t}\begin{bmatrix} 1 & r \\ r & 1 \end{bmatrix}\begin{bmatrix} e^{ik_0 n_1 h} & 0 \\ 0 & e^{-ik_0 n_1 h} \end{bmatrix} \qquad (3)$$

where, $n_1$ and $n_2$ are refractive indices of the first and second layers, $k_0$ is free space wave number, and $h$ is the first layer thickness of plasma. For $x$ plasma layers, the propagation matrix becomes

$$M = M_{x-1} \frac{1}{t}\begin{bmatrix} 1 & r \\ r & 1 \end{bmatrix}\begin{bmatrix} e^{ik_0 n_x h} & 0 \\ 0 & e^{-ik_0 n_x h} \end{bmatrix} \qquad (4)$$

The transmittance $T$ and reflectance $R$ for normal incidence are calculated as $T = |1/M_{11}|^2$ and $R = |M_{21}/M_{11}|^2$, respectively.

*C. Plasma Arrangement effect*

In real-world scenarios, plasma distribution within the sheath is highly inhomogeneous and stochastic, influenced by factors such as spacecraft geometry, attitude, velocity, and surface roughness [26, 27]. Vertical plasma concentration measurements from the sheath exterior to the spacecraft surface often deviate from the idealized 'Low-High-Low' (L-H-L) distribution, as observed in various experiments and simulations [20, 28].

To investigate the impact of plasma arrangement on channel propagation, we compare the typical 'L-H-L' sheath arrangement with a center-inverted 'High-Low-High' (H-L-H) configuration, as illustrated in Fig. 3. We maintain consistent average concentration and length across both arrangements, isolating the concentration distribution as the sole variable. Each layer is set to a uniform thickness of 0.475 cm, matching the ICP model's plasma layer thickness. We perform frequency sweep simulations from 30 GHz to 300 GHz for both arrangements. Due to negligible differences in transmittance above 200 GHz, we focus on the lower frequency range, as shown in Fig. 3(a). Both arrangements exhibit an identical cutoff frequency (~89.9 GHz), corresponding to the maximum electron density of $10^{14} cm^{-3}$. This suggests that the cutoff frequency is determined by the highest-density plasma layer, regardless of the overall arrangement. Besides, the 'H-L-H' arrangement displays more pronounced fluctuations in transmittance, attributed to abrupt changes in electron density. This leads to increased internal reflections at layer interfaces due to larger impedance mismatches and enhanced channel scattering [38]. The phase shift analysis (inset of Fig. 3(a)) confirms this, showing a higher accumulated phase shift for the 'H-L-H' arrangement compared to the 'L-H-L' configuration.

To examine the effect of incident angles on channel transmission [39, 40], we simulate various incidence angles at 150 GHz, a frequency chosen for its sensitivity to plasma layers while exceeding the plasma frequency. Fig. 3(b) shows that both arrangements yield an identical total reflection angle of ~53°, again attributed to the maximum-density plasma layer. The 'H-L-H' arrangement consistently introduces higher transmittance fluctuations across different incident angles.

Our theoretical calculations based on Eq. (4), presented in Fig. 3(c), closely align with the simulation results in Fig. 3(a) for both arrangements. This agreement validates our model's reliability and suggests its applicability for predicting wireless channel performance through various flat plasma layer arrangements, potentially reducing the need for extensive simulations in future scenarios.

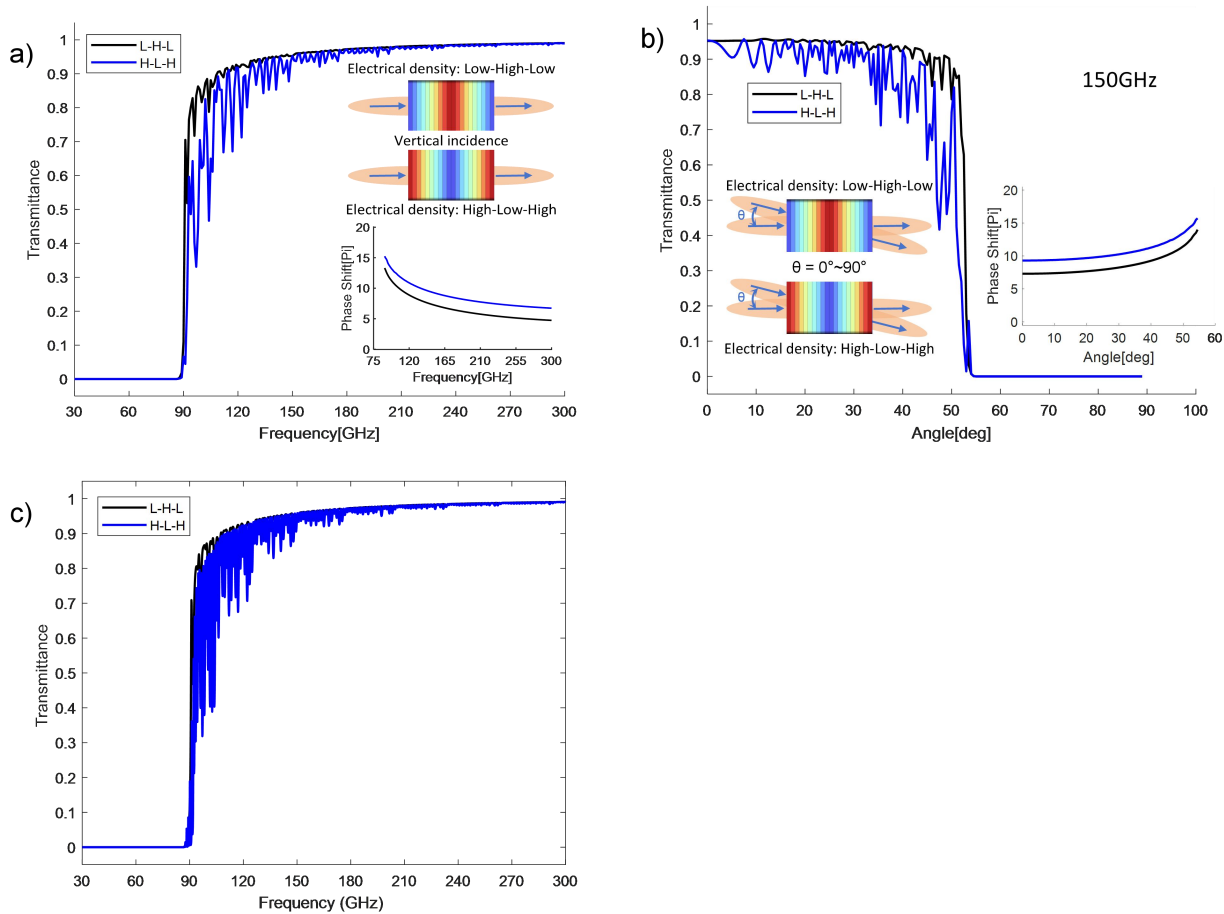

**Fig. 3.** Transmission characteristics of MMW and sub-THz channels through 'L-H-L' and 'H-L-H' plasma layer arrangements. a) Transmittance vs. operation frequency with inset showing corresponding phase shifts by FEA. b) Transmittance vs. incident angle with inset showing corresponding phase shifts by FEA. c) Theoretical prediction of transmittance by TMM.

## III. CHARACTERISTICS OF PLASMA LAYER RIPPLES

The plasma sheath formation during reentry is primarily caused by air ionization due to shock layer effects and viscous heating. Hypersonic aircraft movement compresses and heats the surrounding air, creating a high-temperature, high-pressure environment conducive to molecular ionization. Boundary layer transition and turbulence during flight induce fluctuations in the surrounding gas, resulting in a inhomogeneous plasma sheath surface [28, 29]. While the plasma sheath can mitigate flow separation and control boundary layer transition, it also undergoes significant self-induced structural changes. The interaction between the plasma and the surrounding flow creates a feedback loop, resulting in dynamic reshaping of the sheath. This leads to non-uniform structures such as ripples and density variations. These plasma-induced deformations significantly alter the overall shape and structure of the sheath, directly impacting electromagnetic wave propagation and communication channel performance in hypersonic flight conditions [25]. Simulation data reveals that under specific flight conditions (e.g., specific angle, altitude, aerodynamic configuration), electron density in certain regions exhibits ripple-like distributions. These ripple structures can potentially cover over 50% of the aircraft's surface, depending on factors such as aerodynamic shape and flight angle [26-29]. To investigate these phenomena, we developed a model simulating small-scale ripple fluctuations in the plasma sheath, as shown in Fig. 4. Our model features a sinusoidal ripple structure through which electromagnetic waves propagate at various frequencies and incident angles. To optimize computational efficiency, we simplify the plasma sheath to nine layers, each 0.125 cm thick, approximating the concentration stratification within the sheath. We employ an 'L-H-L' arrangement using plasma data from the ICP model, with five representative plasma concentrations ranging from $1\times10^{13}$ $cm^{-3}$ to $9\times10^{13}$ $cm^{-3}$, increasing in increments of $2\times10^{13}$ $cm^{-3}$.

Using periodic ripple boundary conditions, we simulate propagation through a single period. The model incorporates three amplitudes (4, 6, and 8 mm) and three periods (6, 8, and 10 mm) for the waveform, based on existing plasma sheath simulation studies [20, 28, 29]. For comparative analysis, Fig. 4(a) presents flat layers without ripples. Fig. 4(b-d) illustrate the effects of varying ripple amplitude (with a fixed 10 mm period), while Fig. 4(f-g) show the impact of different periods (with a fixed 6 mm amplitude). We maintain normal incidence onto the sheath's ripple peaks, as depicted in Fig. 4(e), to facilitate analysis of propagation properties under different ripple configurations. Simulation results demonstrate significant changes in channel performance after passing through rippled plasma layers. These effects are particularly pronounced in random ripple structures (Fig. 4(h)), where transmission paths become chaotic. This randomness leads to substantial unpredictable changes in channel performance, complicating the design and optimization of systems relying on precise channel transmission through plasma layers.

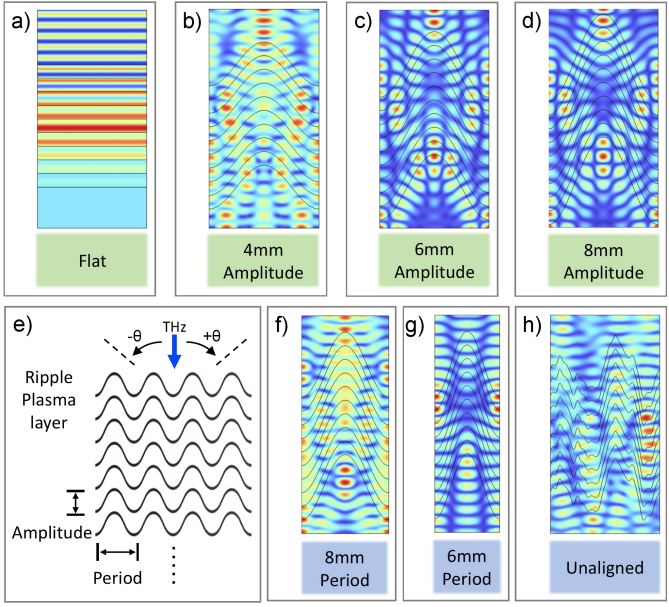

**Fig. 4.** Channel propagation simulations through 9 plasma layers: a) Flat structure; Ripple structures with amplitudes of b) 4 mm, c) 6 mm, and d) 8 mm at a 10 mm spatial period; e) Schematic of normal incidence on the plasma sheath model; Ripple structures with periods of f) 8 mm, g) 6 mm, and h) random values at an 8 mm amplitude.

## A. Ripple Period

The results, presented in Fig. 5, reveal several key findings. Fig. 5(a) demonstrates that the ripple structure reduces the cutoff frequency to approximately 50 GHz. This effect is likely due to spatial variations in plasma layer thickness created by the ripple structure. In thinner regions, waves encounter less plasma, resulting in a more gradual transition. This phenomenon is further corroborated in subsequent analyses of ripple amplitude variations. The ripple structure also introduces significant fluctuations in transmittance, exhibiting frequency-selective behavior. This can be attributed to interference from the periodic ripple structures, potentially degrading signal quality through noise and distortion. The effect is more pronounced for smaller periods due to greater structural fluctuations. While this behavior may be detrimental in some applications, it could prove beneficial in others, such as filter design [1, 2].

To examine frequency band dispersion as the channel propagates through plasma layers, we analyze the Group Velocity Dispersion (GVD) by calculating the second derivative of the wave vector $k$ with respect to angular frequency $\omega$ [41]. Fig. 5(b) clearly shows that GVD for ripple plasma layers is significantly larger than for flat layers (where GVD = 0) and increases for smaller ripple periods. This indicates that different frequency components experience varying delays, leading to more substantial temporal spreading as the channel propagates through ripple plasma layers. Such spreading can cause inter-symbol interference (ISI) and signal distortion, complicating signal decoding.

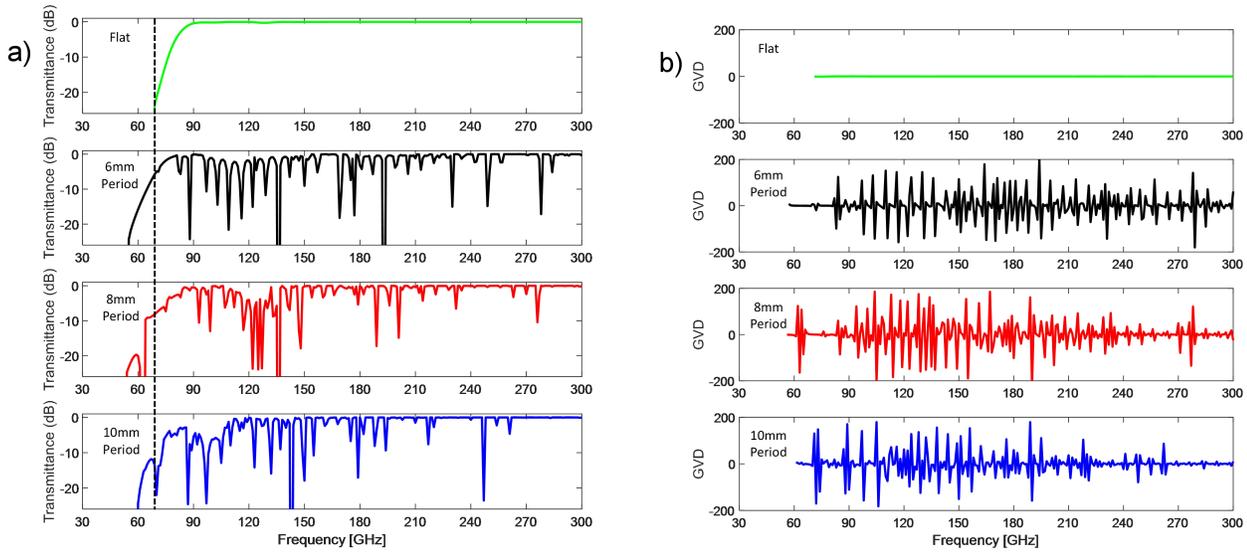

**Fig. 5.** Transmission and GVD performance for a 6 mm amplitude with three different periods: a) Transmittance vs. frequency (30-300 GHz) at constant incident angle; b) GVD vs. frequency (30-300 GHz) at constant incident angle.

We further investigate the effect of incident angle on transmission using a 150 GHz frequency. Fig. 6(a) shows that ripple layers, regardless of period, exhibit the same critical angle (of total reflection) at 65º, which is greater than the 63º observed for flat layers. This is due to the ripple structure altering the relative incident angle of the channel, as previously noted in Fig. 3(b). The ripple structure introduces angular selectivity in the plasma sheath's reflection and transmission properties. However, it also restricts available transmission angles, reducing the angular transparency window. Increasing the ripple period further narrows this window and diminishes transmission loss peaks. These effects could limit channel direction flexibility, complicate antenna alignment, and may lead to reduced coverage area and increased interference and multipath effects. The GVD performance in Fig. 6(b) further confirms these observations.

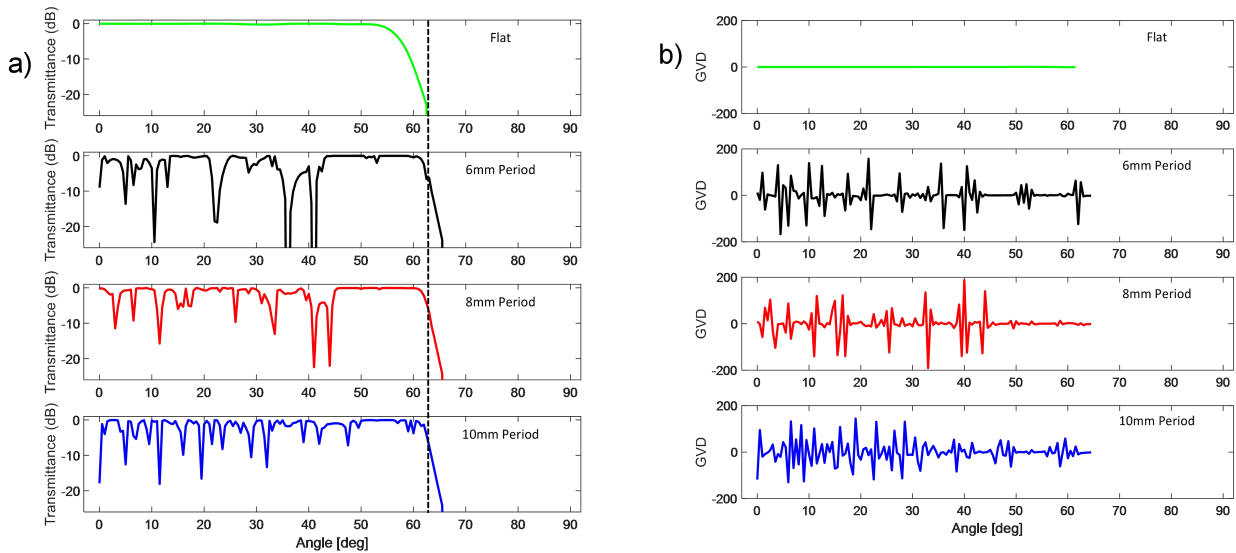

**Fig. 6.** Transmission and GVD performance for three different periods with a 6 mm amplitude: a) Transmittance vs. incident angle (0º - 90º); b) GVD vs. incident angle (0º - 90º). .

## B. Ripple Amplitude

We conduct simulations with varying ripple amplitudes, for further elucidating the impact of ripple structure on channel transmission. As illustrated in Fig. 7(a), increasing the amplitude from 4 mm to 6 mm and 8 mm progressively reduces the cutoff frequency, consistent with our observations in Fig. 5(a). This phenomenon can be attributed to the fact that larger ripple amplitudes create more significant variations in plasma layer thickness, resulting in a softer cutoff with a more gradual transition. The increase in ripple amplitude also introduces more severe fluctuations in transmission characteristics. These fluctuations are a direct result of the variations in refractive index within the plasma layers caused by the ripple structure. Consequently, this leads to enhanced scattering and interference effects, culminating in more pronounced dispersion. Notably, the 8 mm amplitude case exhibits the most severe degradation in transmission, which aligns with the GVD performance shown in Fig. 7(b). This increased dispersion manifests as greater temporal spreading and phase shifts in the transmitted signal, significantly impacting overall transmission characteristics.

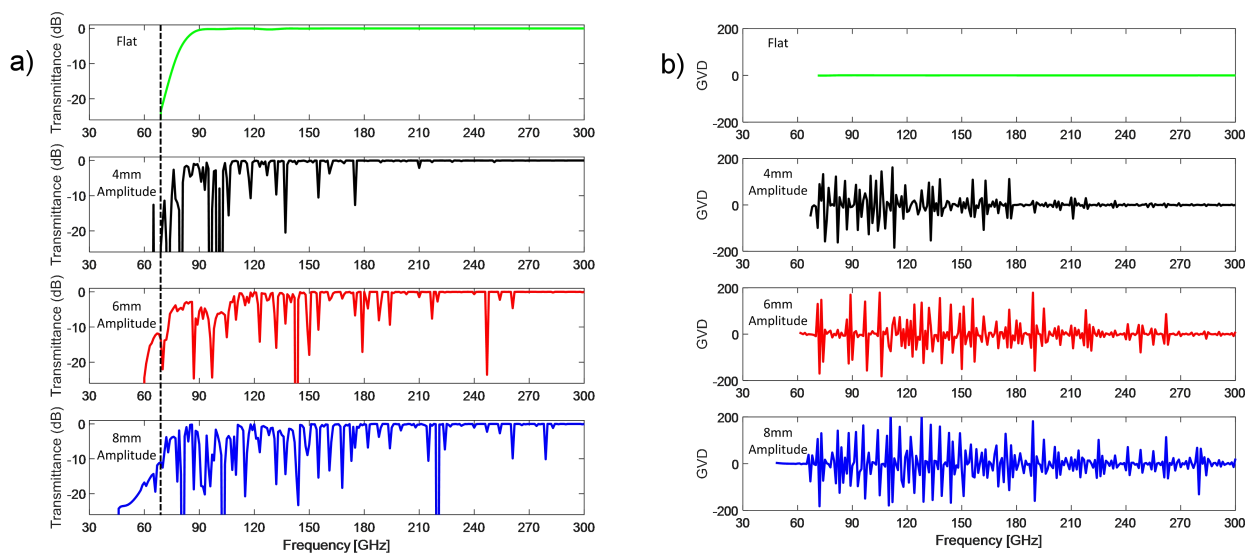

**Fig. 7.** Transmission and GVD performance for three ripple amplitudes with a 10 mm period: a) Transmittance vs. frequency (30-300 GHz) at constant incident angle; b) GVD vs. frequency (30-300 GHz) at constant incident angle.

To further investigate the angular dependence of transmission, we maintain a constant operating frequency of 150 GHz and simulated irradiation on plasma layers at various incident angles. The results, depicted in Fig. 8(a) and (b), illustrate the transmission and GVD performance, respectively. A key observation is that the presence of ripples increases the critical angle (angle of total reflection). Specifically, the critical angle changes from 63º in flat layers to 66º (4 mm amplitude), 67º (6 mm amplitude), and 68º (8 mm amplitude) in rippled layers. We posit that this phenomenon arises due to the increase in amplitude affecting the beam's relative incident angle. Under a fixed incident angle to the whole plasma layer structure, a larger amplitude increases the range of relative incident angles experienced by the channel beam as it traverses the rippled surface.

Based on the above observations, we can say that the periodic ripple plasma structures effectively modulate the wireless signals as they propagate through or are reflected by them. This modulation effect has several important implications for wireless communication systems operating in environments with plasma sheaths, such as frequency-dependent transmission, angular sensitivity, dispersion management, *etc*.

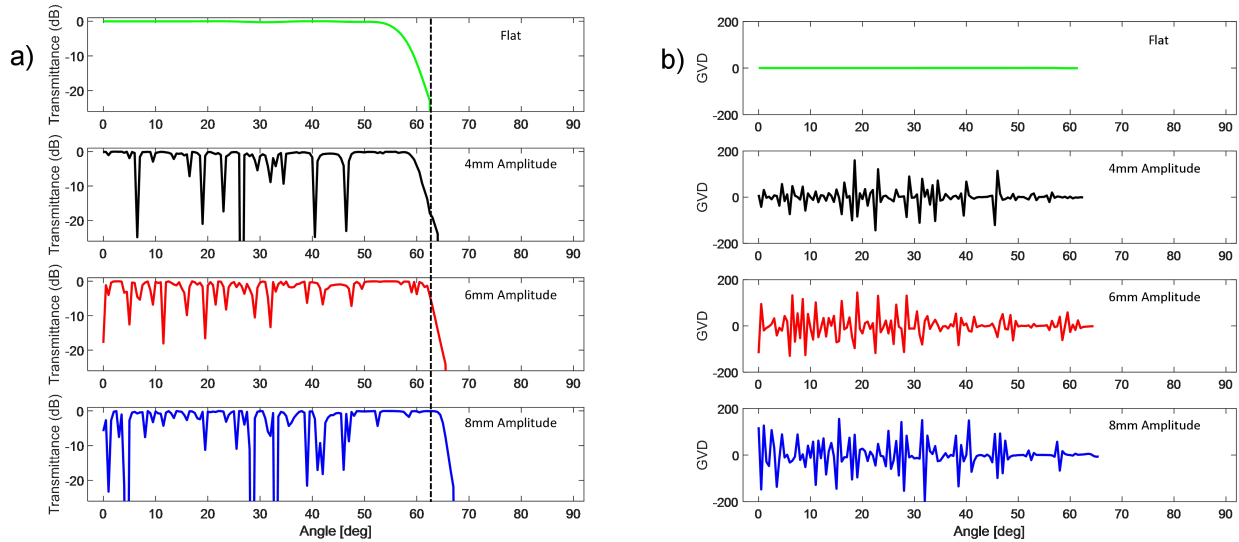

**Fig. 8.** Transmission and GVD performance for a 10 mm period with three amplitudes and flat model: a) Transmittance vs. incident angle (0º - 90º); b) GVD vs. incident angle (0º - 90º).

*C. Random Ripple*

In realistic scenarios, the plasma layers surrounding hypersonic aircraft often exhibit random ripple structures, as illustrated in Fig. 4(h), not the periodic distributions previously discussed. Our simulation of these random ripple layers, presented in Fig. 9(a), reveals that they do not cause a significant shift in the cutoff frequency compared to flat layers, unlike the observations made for periodic ripples in Fig. 7(a). This finding suggests that while random ripples create more complex variations in the refractive index, their overall impact on the cutoff frequency is averaged out, resulting in a less pronounced effect on this particular parameter. However, it's crucial to note that random ripple structures do introduce more severe fluctuations in transmission characteristics. These fluctuations lead to increased scattering and interference effects, potentially causing more disruption to channel transmission than periodic ripples. The intensified scattering and interference of electromagnetic waves in random structures could pose significant challenges for maintaining stable and reliable communication channels.

Examining the angular transmission characteristics through random ripple layers, as shown in Fig. 9(b), we observe similarities with the periodic layers depicted in Fig. 8(a), with one notable exception: the critical angle of total reflection. Interestingly, for random ripples, this critical angle matches that of flat layers. This observation has important implications for real-world applications. In scenarios where plasma sheaths exhibit random ripples, which is likely in actual hypersonic flight conditions, the cutoff frequency and critical angle of total reflection may be more predictable and closer to those of an idealized flat sheath.

However, it's important to emphasize that while the cutoff frequency and critical angle may be more predictable, other channel propagation characteristics, such as localized scattering effects, multipath propagation, and small-scale fading, are still significantly influenced by random ripples. Therefore, while the random ripple structure provides some predictability in terms of cutoff frequency and critical angle, it introduces new challenges in other aspects of channel characterization and system design.

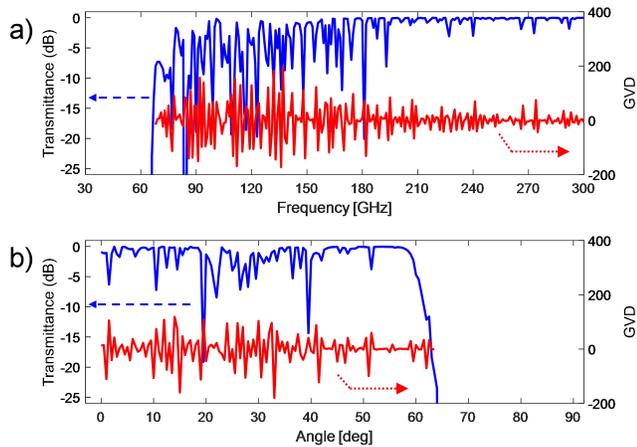

**Fig. 9.** Transmission and GVD performance through random ripple layers as a function of: a) operating frequency and b) incident angle.

## IV. Conclusion

In this work, we conduct a comprehensive investigation into the transmission characteristics of MMW and sub-THz channels through plasma sheaths with various structural configurations. We develop and validate a model that combines ICP data with TMM method to analyze the propagation of wireless channel through both periodic and random ripple plasma layers. We compare 'L-H-L' and 'H-L-H' plasma density arrangements, demonstrating that while the cutoff frequency remains consistent, the 'H-L-H' configuration introduces more severe fluctuations in transmittance and phase shift. We also find that both periodic and random ripple structures in plasma layers significantly affect channel performance. Periodic ripples reduce the cutoff frequency and introduce frequency-selective behavior, while random ripples maintain a cutoff frequency similar to flat layers but cause more unpredictable transmission characteristics. Smaller ripple periods and larger amplitudes generally lead to more pronounced effects on channel transmission, including increased GVD and changes in the critical angle of total reflection. Finally, while random ripples create more complex refractive index variations, their effect on cutoff frequency and critical angle is less pronounced than periodic ripples, potentially simplifying some aspects of system design.

These findings have significant implications for the design and optimization of communication systems for hypersonic vehicles. The complex interactions between wireless channels and plasma sheaths, particularly those with ripple structures, necessitate careful consideration in system design. Future communication systems for hypersonic applications may need to incorporate adaptive technologies capable of adjusting to varying plasma conditions in real-time.

### Acknowledgment

This work was supported in part by the National Natural Science Foundation of China (62071046), the Science and Technology Innovation Program of Beijing Institute of Technology (2022CX01023), the Graduate Innovative Practice Project of Tangshan Research Institute, BIT (TSDZXX202201), the Fundamental Research Funds for the Central Universities (2024CX06099) and the Talent Support Program of Beijing Institute of Technology "Special Young Scholars" (3050011182153).